\documentclass[prd,showpacs,amsmath,showkeys,twocolumn,floatfix]{revtex4} 
\usepackage{epsfig,dcolumn}
\usepackage{graphicx}
\usepackage{graphicx}
\usepackage{bm}

\def\qq{{q{\bar q}}}
\def\x{{\bf x}}
\def\y{{\bf y}}
\def\z{{\bf z}}
\def\k{{\bf k}}
\def\q{{\bf q}}
\def\p{{\bf p}}

\def\A{{\bf A}}

\def\lsim{\mathrel{\rlap{\lower4pt\hbox{\hskip1pt$\sim$}}
    \raise1pt\hbox{$<$}}}
\def\gsim{\mathrel{\rlap{\lower4pt\hbox{\hskip1pt$\sim$}}
    \raise1pt\hbox{$>$}}}
\begin{document}


\title{ Energy spectrum of the low-lying gluon excitations in the Coulomb gauge}

\author{ Adam P. Szczepaniak and Pawel Krupinski}
\affiliation{ Physics Department and Nuclear Theory Center \\
Indiana University, Bloomington, Indiana 47405 }

\date{\today}

\begin{abstract}
We compute the energy spectrum of low-lying gluonic excitations in the presence of static quark-antiquark sources using Coulomb gauge and the quasi-particle representation.
Within the valence sector of the Fock space we reproduce both, the overall normalization and the ordering of the spin-parity multiplets. We discus how the interactions induced by the non-abelian Coulomb kernel are central in to fine structure of the spectrum. 
  \end{abstract}

\pacs{11.10Ef, 12.38.Aw, 12.38.Cy, 12.38.Lg}





\maketitle
\section{Introduction} 
Lattice simulations of the energy of static quark-quark or quark-antiquark ($Q{\bar Q}$) systems 
 have been studied for various configurations of the gluonic field~\cite{Juge:1997nc,Juge:2002br,Bali:2005fu,Takahashi:2004rw}. They effectively represent 
 the spectrum of gluonic 
 excitations in the presence of sources. The results of simulations indicate that the spectrum is quite complex with an unusual ordering of levels and unexpected behavior for large separations between  the sources.  In the following we will specialize to the spectrum in the presence of $Q{\bar Q}$ sources and in the absence of light quark pairs. The wave function of the gluon field  can be characterized according to its behavior under the symmetries of the $Q{\bar Q}$ system. 
 These include rotation around the $Q{\bar Q}$ axis, hence after chosen to be in the $z$ direction. The corresponding conserved quantum number, $\Lambda$ represents projection of the total gluon angular momentum on the $Q{\bar Q}$ axis and is conventionally denoted by $\Lambda=\Sigma, \Pi, \Delta,\cdots$ corresponding to projections, $\Lambda=0,1,2,\cdots$, respectively. Other symmetries are  the combined product of parity and charge conjugation, $PC$, with eigenvalues denoted by $g$ and $u$ for $PC=+1$ and $-1$, respectively and  reflection in the $x-z$ plane, $Y=\pm 1$. The representation space of QCD eigenstates in the presence of static $Q{\bar Q}$ sources is then given by 
  states labeled as
   $|R,\Lambda^{PC}_Y\rangle$ where $R$ is the distance between the sources. The corresponding energies will be denoted by $V_{\Lambda^{PC}_Y}(R)$. 

From lattice simulations the following characteristics of the the spectrum emerge. As expected, the ground state has $\Sigma^+_g$  quantum numbers and as a function of $R$ is well described by the "Coulomb+linear", Cornell-type potential. The first excited state has one unit of gluon angular momentum 
along the $Q{\bar Q}$ axis and {\it negative} $PC$. The state with $\Lambda=1$ and positive $PC$ has energy higher by $~400\mbox{ MeV}$ at $R=1\mbox{ fm}$. This fact alone can be used to discriminate between various pictures describing the dynamics of gluonic modes. For example, if the gluon field  is thought of as a localized quasi-particle that interacts with the quark sources~\cite{Horn:1977rq}  in a way that satisfies Casimir scaling~\cite{Bali:2000un}, then it would be expected that the $S$ wave, spherically symmetric gluon  wave function has lower energy than the $P$-wave with one unit of orbital angular momentum. Since gluons have internal spin-parity quantum numbers $J^{PC} = 1^{--}$ this would result in $\Pi_{g}$ ($PC=1$) state having lower energy than the $\Pi_u$ ($PC=-1$) contradicting lattice results. 
A similar reversed ordering is observed in higher excitations as well.  The $\Delta_g$ state has lower energy than $\Delta_u$ which in the constituent picture means that the gluon state with two units of orbital angular momentum has energy lower than the state with one unit. The same is true for excited $\Sigma$ states. The $\Sigma_u^-$ state which in the constituent picture has one unit of orbital angular momentum has lower energy than the first excited $\Sigma'^+_g$ with vanishing  orbital angular momentum. Higher in the spectrum  the $\Sigma_g^-$ corresponding to two units of orbital angular momentum  has lower energy than $\Sigma^+_u$ with one  unit of orbital angular momentum, albeit the splittings between these higher excitations are smaller than for the $\Pi$ states. These inconsistencies between lattice results and the constituent picture have been noticed in ~\cite{Swanson:1998kx}. The bag 
 model~\cite{Hasenfratz:1980jv} seems to be doing better in this respect. The boundary conditions on the gluon field, which set it free inside the bag cavity make the TE mode with pseudo-vector, $J^{PC} = 1^{+-}$, quantum numbers to have lower energy than the TM mode with $J^{PC} = 1^{--}$. This in 
  turn leads to  the  energy of $\Pi_u$  state to be lower than for the $\Pi_g$ state~\cite{Juge:1997nd}. Finally in ~\cite{Swanson:1998kx} it was observed that the non-relativistic flux tube model~\cite{Isgur:1984bm}  also predicts the state with $\Lambda=1$ and $PC=+1$ quantum numbers to be the first excited state.  This is because in the flux tube model the gluon degrees of freedom moving in a plane transverse to the $Q{\bar Q}$ axis have negative parity but, unlike the vector potential have positive internal charge conjugation. 
 Thus both the bag model and the flux tube model give the right ordering of the spectrum of low lying gluonic excitations, albeit  for different reasons. 
 The constituent gluon picture of ~\cite{Swanson:1998kx} was based on the mean field representation of the Coulomb gauge QCD, however it did not take into account the interactions emerging from the 
 non-abelian Coulomb potential. In ~\cite{Szczepaniak:2005xi} it was shown how such interactions affect the $\Sigma_g^+$ potentials. Here we extend that analysis to  gluonic excitations with other symmetries. The main finding is that  the non-abelian Coulomb potential is responsible for reversing the naive ordering expected from two body quark-gluon interactions. We also find that the overall scale of gluonic excitations in the mean-field quasi-particle picture 
  is somewhat higher but consistent with the lattice results  thus making the constituent gluon model a viable representation of the low energy gluon dynamics. This representation also explains the 
 degeneracies in the spectrum seen in the lattice data at  small $Q{\bar Q}$ separations. 
     Such degeneracies are expected as the system becomes more spherically symmetric. At large $Q{\bar Q}$ separations the gluonic wave function is expected to be qualitatively different from that of a single or a few quasi-particle state~\cite{Thorn:1979gu,Greensite:2001nx}. As separation between color sources increases the mean field Coulomb interaction is expected to rise more rapidly than the true energy of the system~\cite{Greensite:2004ke,Zwanziger:2002sh,Nakagawa:2006fk}. Thus as the separation increases states with a large number of gluons separated, on average,  by a  small  fraction of the $Q{\bar Q}$ distance $R$  are expected to have lower energy than states with a small number of gluons separated by a distance of the order of the $Q{\bar Q}$ separation. At what $Q{\bar Q}$ distances 
      the transition from the constituent, few gluon picture to the flux tube or string like picture 
       takes place is however still an open question. Analysis of lattice results for the splitting between  gluonic levels at large $R$ does not conclusively favor the string like picture even for $Q{\bar Q}$ separations as large as a few fermi~\cite{Juge:2004xr,colin}. In~\cite{Szczepaniak:2005xi} we have shown that if the mean field Coulomb energy also rises linearly with the $R$, as indicated by lattice computations~\cite{Greensite:2004ke} then it is difficult to generate the constituent string and the state is dominated by valence constituent gluons. This is  because the Coulomb energy of a state of $n_g$ gluons separated from each other by a distance $O(R/n)$  does not depend on the number of gluons~\cite{Jeff},  thus an increase in the average number of gluons can only originate from mixing. However 
         as $R$ increases it  turns out that mixing between states with a different number of gluons  
          is  much smaller than the diagonal Coulomb energy~\cite{Szczepaniak:2005xi}. Lattice results of 
       ~\cite{Greensite:2004ke} show that the Coulomb energy rises linearly with the $Q{\bar Q}$ separation with a string tension which is roughly three times bigger than that of the true energy, however, since  the lattice $Q{\bar Q}$ state used to compute the Coulomb energy is not the same at the mean field state which defines the quasi particle gluon basis it is possible that the  Coulomb energy of the latter grows with $R$ faster than linearly, {\it i.e} as $R^\gamma$ with $\gamma>1$. In  this case the average number of gluons at fixed $R$  would roughly correspond to the minimum with respect to $n_g$ of 
       \begin{equation}
       E_{n_g}(R) = n_g m_g  + n_g \left( {R \over {n_g}} \right)^\gamma, 
       \end{equation}
 leading to $n_g\sim R$ and energy rising linearly with $R$. Here $m_g$ represents the average kinetic energy of a quasi-gluon. It is thus possible that there is a smooth connection between the quasi-particle and flux tube representation, but we leave a more  quantitative  description for the future. 
  In the following we  explore the fine structure of the Coulomb interaction  and its role in ordering the spectrum of low-lying gluonic excitations. We will therefore restrict the quasi-gluon Fock space to a the single particle sector  and fit the expectation value of the mean field Coulomb interaction to the ground state $Q{\bar Q}$ energy~\cite{Szczepaniak:2005xi,Szczepaniak:2003ve,Szczepaniak:2001rg}. 
       
       The paper is organized as follows. In the next section we summarize the basics of the mean field Coulomb gauge approach and describe the relevant interactions. In Section III we present out results for the spectrum. A summary and outlook are given in Section IV.

  \section{Coulomb gauge QCD Hamiltonian in the quasi-particle representation} 

The derivation of QCD in the Coulomb gauge is given in~\cite{Christ:1980ku}. The canonical approach is to start from $A^{0,a}(\x) = 0$ the Weyl gauge and using the residual gauge freedom perform a coordinate transformation  to coordinates ${\bf A}^a(\x)$  constrained to satisfy   $\bm{\nabla}\cdot {\bf A}^a(\x) = 0$  and ${{\bf N_C^2-1}}$ phases $\phi^a(\x)$, $a=1,\cdots, N_C^2-1$. In the new coordinates  Gauss's law can be used to eliminate the dependence on the gauge phases and in the Shr{\"o}dinger representation the  QCD spectrum is formally obtained by solving 
\begin{equation}
H\left[ {\bf A}^a(\x), \bm{\Pi}^a(\x) \right] \Psi_n[{\bf A}^a]  = E_n \Psi_n[{\bf A}^a],  
\end{equation} 
with the canonical momenta,  $\bm{\Pi}^a = -i\partial /\partial{\bf A}^a(\x)$, 
 satisfying  $[\Pi^a_i(\x),A_j^b(\y)] = -i \delta_{ab}\delta^{ij}_T(\bm{\nabla})\delta^3(\x-\y)$ where $\delta^{ij}_T(\bm{\nabla}) = \delta^{ij} - \nabla^i\nabla^j/\bm{\nabla}^2$. 
  The coordinate transformation from the Weyl gauge to the Coulomb gauge is nonlinear and leads to the Faddeev-Popov (FP) determinant in the space of the field  configurations, ${\cal J} = Det(1-\lambda)$, 
  where, $(1-\lambda)(\x,a;\y,b) =\delta_{ab}\delta^3(\x-\y) - (g/4\pi) f_{acb}  \bm{\nabla}_y (1/|\x - \y|) \A^c(\y)$.  More discussion of the topological properties of the fundamental domain of the gauge variables can be found in ~\cite{vanBaal:1997gu}. The role of FP determinant has been investigated
   in~\cite{Szczepaniak:2005xi,Reinhardt:2004mm,Feuchter:2004mk} where it  was found that it can be effectively absorbed into the parametrization of the vacuum wave functional. In the following we will thus set ${\cal J}=1$ and use the ansatz for the variational (unnormalized) vacuum wave functional of the form, ~\cite{Szczepaniak:2001rg}

\begin{equation}
\langle  {\bf A}| 0\rangle =\Psi_0[{\bf A}] =  \exp\left( - \int {{d\k} \over {(2\pi)^3} } {\bf A}^a(\k) \omega(k) {\bf A}^a(-\k) \right),
\end{equation}
where ${\bf A}^a(\k) = \int d\x \exp(-i\k\cdot\x) {\bf A}^a(\x)$ is the Fourier transform for the coordinate space. 
  
  The variational parameter $\omega(k)$ is obtained by solving the Dyson (gap) equation arising from minimizing the vacuum expectation value ({\it vev}) of the Hamiltonian   $\partial\langle 0|H|0\rangle/\partial \omega(k) = 0$. The solution is well approximated by $\omega(k) = m_g$ for $k=|\k|\le m_g$ and $\omega(k) = k$ for $k>m_g$ with $m_g = 600\mbox{ MeV}$~\cite{Szczepaniak:2001rg}. 
  In computing the {\it vev} of the Hamiltonian the Coulomb energy  
  \begin{equation}
H_{C} = {1\over 2} \int d\x d\y \rho_g^a(\x) K[\A](\x,a;\y,b) \rho^b_g(\y),
\end{equation}
 contributes the energy of interaction between color charges 
  $\rho_g = -f_{abc} \bm{\Pi}^b(\x)\cdot \A^c(\x)$.  In Coulomb kernel, $K[A]$  appearing in $H_C$ 
   the self-interactions between the transverse gluons 
   \begin{equation} 
 K[\A](\x,a;\y,b) = {{g^2}\over {4\pi}}  \int d\z {{  (1 - \lambda)^{-2}(\x,a;\z,b) } \over {|\z - \y|}} ,
  \end{equation}
  are evaluated to leading order in $N_C$. This enables to express the {\it vev} of $K$ in terms of a set of two coupled Dyson equations whose solution can be written as 
  \begin{equation}
  \langle 0|K[\A](\x,a;\y,b) |0\rangle/\langle0|0\rangle = -\delta_{ab} V_C(|\x - \y).
  \end{equation}
  The expectation value of the kernel $V_C$ is renormalized in such a way that the expectation value of the Coulomb interaction between $Q{\bar Q}$ sources in the mean field gluon vacuum reproduces the lowest energy of the $Q{\bar Q}$  state~\cite{Szczepaniak:2001rg,Szczepaniak:2005xi}.

The complete spectrum of gluon state can be obtained by successive application of quasi-particle, gluon creation operators, $\alpha^a(\k,\lambda)$, defined with respect to the mean field vacuum, through the BCS transformation

\begin{eqnarray}
\A^a(\x) & =  & \int {{d\k} \over {(2\pi)^3}} {1\over {\sqrt{2\omega(k)}}} \left[ \alpha^a(\k,\lambda) \bm{\epsilon}(\k,\lambda) \right. \nonumber \\
 & &  \left. + \alpha^{a,\dag}(-\k,\lambda) \bm{\epsilon}(-\k,\lambda) \right ] e^{i\k\cdot\x}, \nonumber \\
 \bm{\Pi}^a(\x) & = & -i \int {{d\k} \over {(2\pi)^3}}  \sqrt{ {\omega(k)}  \over 2} 
\left[ \alpha^a(\k,\lambda) \bm{\epsilon}(\k,\lambda) \right. \nonumber \\
 & &  \left. - \alpha^{a,\dag}(-\k,\lambda) \bm{\epsilon}(-\k,\lambda) \right ] e^{i\k\cdot\x}. \nonumber \\
 \end{eqnarray}
 Here $\bm{\epsilon}$ represent  helicity vectors with $\lambda =\pm 1$.  When describing the spectrum of gluons in the presence of $Q{\bar Q}$ sources, we will truncate the quasi-particle gluon Fock space to contain a single quasi-gluon, {\it i.e.} the Hamiltonian will be diagonalized in the basis of states spanned by 
 
   \begin{equation}
  |R,\k,\lambda\rangle = {1\over {\sqrt{N_C C_F}}} Q^{\dag}_{{R\over 2}\hat\z}  \alpha^{\dag,a}(\k,\lambda)T^a {\bar Q}^{\dag}_{-{R\over 2} \hat\z} {{|0 \rangle} \over {\langle 0|0\rangle} }. \label{Fs}
  \end{equation}

\section{Spectrum of low-lying gluonic excitations}
 
 From the single quasi-particle state given in Eq.~(\ref{Fs}) the states with good $\Lambda^{Y}_{PC}$ quantum numbers can be constructed 
 
  \begin{eqnarray}
& &  |R,N,\Lambda^Y_{PC} \rangle  =    
   \int {{d\k} \over {(2\pi)^3}}  \sum_{j_g,\xi,\mu,\lambda} 
  \psi^{j_g}_{N}(k) \chi^{\xi}_{\mu\lambda} |R,\k,\lambda\rangle  \nonumber \\ 
&& \times   \sqrt{ {2j_g+1}\over {8\pi}} 
 \left[ D^{j_g*}_{\Lambda\mu}(\hat\k) 
 + \eta_Y  D^{j_g*}_{-\Lambda\mu}(\hat\k)  \right], 
  \label{wfg} 
    \end{eqnarray}
  for $\Lambda \ne 0$ and 
   \begin{eqnarray}
& &  |R,N,0^{PC}_Y \rangle =   \int {{d\k} \over {(2\pi)^3}}    \sum_{j_g,\xi,\mu,\lambda}
 \psi^{j_g}_{N}(k) \chi^{\xi}_{\mu\lambda} |R,\k,\lambda\rangle \nonumber \\
&& \times  \sqrt{ {2j_g+1}\over {4\pi }}  D^{j_g*}_{0\mu}(\hat\k),  
  \label{wfg1} 
    \end{eqnarray}
for $\Lambda=0$.
Here $j_g$ represents the total angular momentum of the quasi-gluon. It is a good quantum number only in the limit $R\to0$, while for finite $R$ states with different values of $j_g$ can mix, although in our numerical computations we have found that of a single $j_g$ state dominates the energy eigenstates for all values of $R$ considered. It is 
 only the projection of the total angular momentum on the $Q{\bar Q}$ axis, $\Lambda$, that is always conserved.  The wave function $\chi^\xi_{\mu\lambda}$ represents the two possibilities for the spin-oribt coupling of given parity.  It is given by
$\delta_{\mu\lambda}/\sqrt{2}$ for $\xi= 1$ ($j_g = L_g \pm 1$) and  $\lambda\delta_{\mu\lambda}/\sqrt{2}$ for $\xi=-1$ ($j_g = L_g$), corresponding to TM (natural parity) and TE (unnatural parity) gluons, respectively. The parity under reflection in the $x-z$ plane, $Y=\pm  1$, for $\Lambda \ne 0$ is determined by $\eta_Y=\pm 1$.  The radial wave functions, $\psi^{j_g}_N(k)$ are 
  labeled by an excitation number $N$ and $j_g$ and are solutions of the Coulomb gauge Hamiltonian projected onto the single quasi-gluon basis 
\begin{equation}
P H  P |R,N, \Lambda^Y_{PC} \rangle = V_{N,\Lambda^{PC}_Y}(R) |R,N,\Lambda^Y_{PC} \rangle. \label{hgbare}
\end{equation}
Here $P$ projects on the $|\qq g\rangle$ states and $V_{N,\Lambda^{PC}_Y}(R)$ are the energies that will be  compared with the lattice spectrum.  The matrix elements of $PHP$ are shown in Fig.~\ref{me} and given explicitly in the Appendix.

 \begin{figure}
\includegraphics[width=3in]{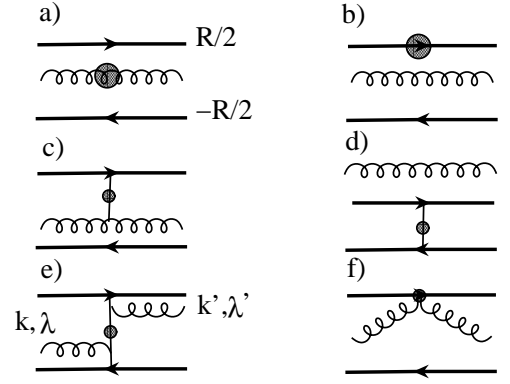}
\caption{\label{me} Matrix elements, $\langle R, \k',\lambda' | H| R,\k, \lambda \rangle$. Diagrams a) and b) represent gluon and quark self energies, respectively. Diagrams c) and d) represent the Coulomb interaction, $V_C$ between the gluon and one of the quarks and between the two quarks, respectively.  In the bottom row, diagrams e) and f)  describe matrix elements of the interaction term resulting from expansion of the Coulomb kernel $K[A]$ in up to one power in  gluon field. }
 \end{figure}

\begin{figure}
\includegraphics[width=2.5in,angle=270]{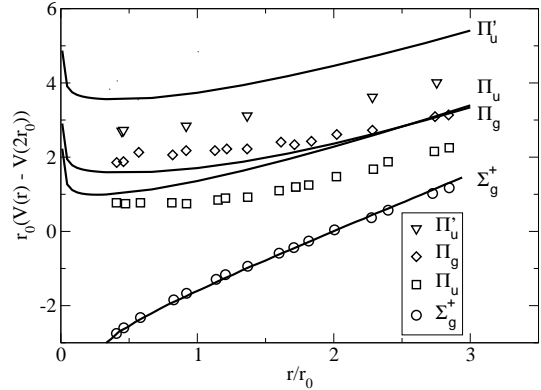}
\caption{\label{2b-p} Spectrum of the $\Lambda=1$, $\Pi$ states compared to the ground state $Q{\bar Q}$ potential {\it without} the three body quark-antiquark gluon interaction corresponding to the diagram $e$ in Fig.~\ref{me}. Lattice results are taken from~\cite{Juge:1997nc}. The energy scale is $r_0 = (400\mbox{ MeV})^{-1}$.}
 \end{figure}

\begin{figure}
\includegraphics[width=2.5in,angle=270]{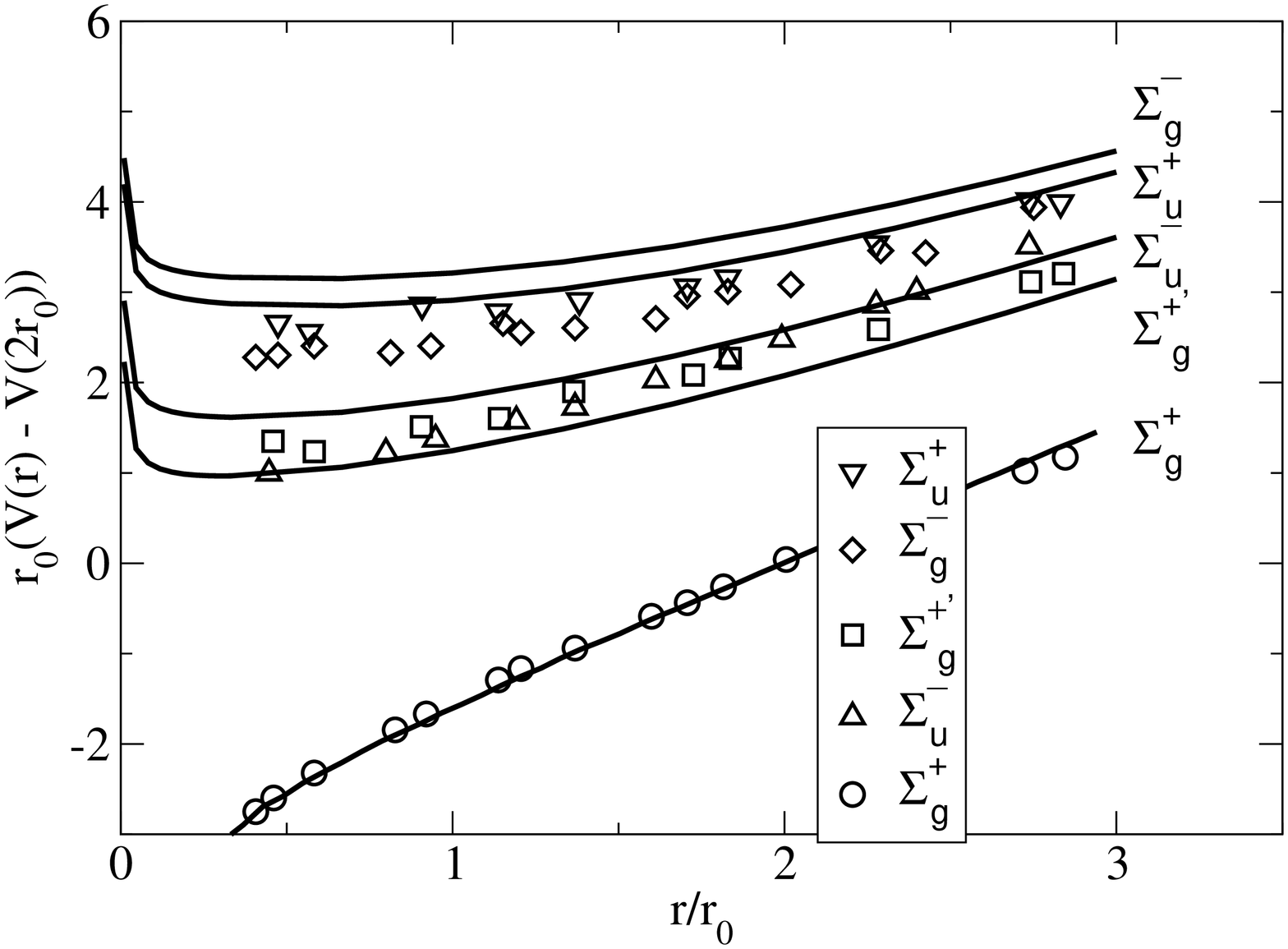}
\caption{\label{2b-s} Same as in Fig.~\ref{2b-p} for the $\Lambda=0$, $\Sigma$ energies. }
\end{figure}

\begin{figure}
\includegraphics[width=2.5in,angle=270]{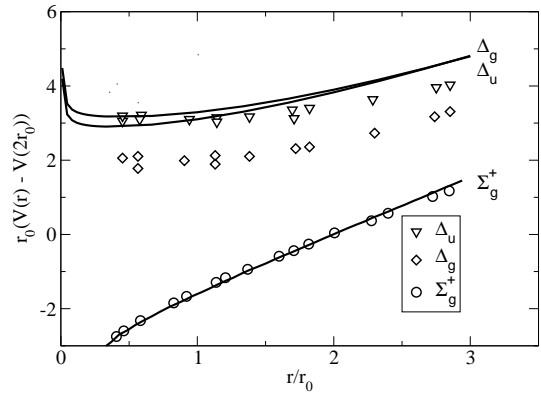}
\caption{\label{2b-d} Same as in Fig.~\ref{2b-p} for the $\Lambda=2$, $\Delta$ energies. }
 \end{figure}

In terms of $\xi$ and $\eta$, the $PC$ and $Y$ quantum numbers  of the gluonic field are given by 
 \begin{equation}
 PC = \xi (-1)^{j_g + 1} = (-1)^{L_g}, \; Y= \left\{ \begin{array}{c} \xi \eta_Y (-1)^\Lambda  \mbox{ for } \Lambda \ne 0 
 \\ \xi \mbox{ for } \Lambda = 0 \end{array} \right.  .  \label{PC} 
  \end{equation} 
For $\Lambda \ne 0$ the two $Y=\pm 1$ states are degenerate. 
For small $Q{\bar Q}$ separations the pattern of the spectrum measured on he lattice can easily be understood  since $j_g$ becomes a good quantum number. In this case the gluon configurations of the eight lowest excitations are given in Table~\ref{T1}.  By setting $R=0$ in the Hamiltonian matrix elements which involve gluon-quark interactions (see Appendix) one finds the following.  The interactions become $\Lambda$ independent. This is expected since in the limit $R\to 0$
 the system does not have a preferred  direction while $\Lambda$ selects one. The angular momentum barrier  pushes states with higher orbital angular momentum, $L_g$  up in energy.
The quark-gluon and antiquark-gluon interactions are attractive, while the quark-antiquark interaction in the color octet channel is repulsive. Thus the Coulomb tail of the quark-antiquark potential will eventually lead to rising energies as $R\to 0$. Outside of this short distance Coulomb region, however  
 for given $j_g$ and $\xi$ we expect states with different $\Lambda$ to be degenerate.  Further, 
  the $\xi=+1$ multiplets (which contain $L_g = j_g-1$) are expected to be lower in energy than the $\xi=-1$  multiplets which have $L_g=j_g$. This is indeed seen in the numerical results
    shown in Figs.~\ref{2b-p},\ref{2b-s},\ref{2b-d}.  Compared to the lattice results, however we see that while the degeneracy  between different $\Lambda$  states within  $\xi$-multiplets indeed occurs the order of the $\xi=-1$ and $\xi=+1$  multiplets is reversed.

\begin{table}
\begin{tabular}{|c|c|c|c|}
\hline
State & $\xi$  & $j_g$ & $L_g$    \\
\hline
\hline
$\Pi_u,\Sigma^-_u$ & $-1$ & $1$ & $1$ \nonumber \\
\hline
$\Pi_u,\Sigma'^+_g$ & $+1$ & $1$ & $0,2$ \nonumber \\
\hline
$\Delta_g,\Sigma^-_g$ & $-1$ & $2$ & $2$ \nonumber \\
\hline
$\Delta_u,\Sigma^+_u$ & $+1$ & $2$ & $1,3$ \nonumber \\
\hline
\end{tabular}
\caption{\label{T1} Spin-orbital wave functions of the lowest single quasi-gluon states.} 
\label{table1}
\end{table}

\begin{figure}
\includegraphics[width=2.5in,angle=270]{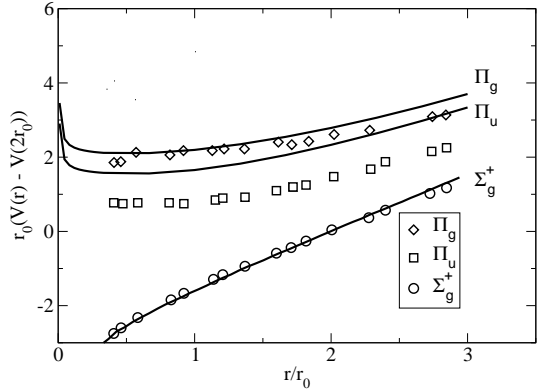}
\caption{\label{3b-p} Spectrum if the $\Lambda=1$, $\Pi$ states compared to the ground state $Q{\bar Q}$ potential using the complete Hamiltonian, which {\it includes} the three body quark-antiquark gluon interactions.}
 \end{figure}

\begin{figure}
\includegraphics[width=2.5in,angle=270]{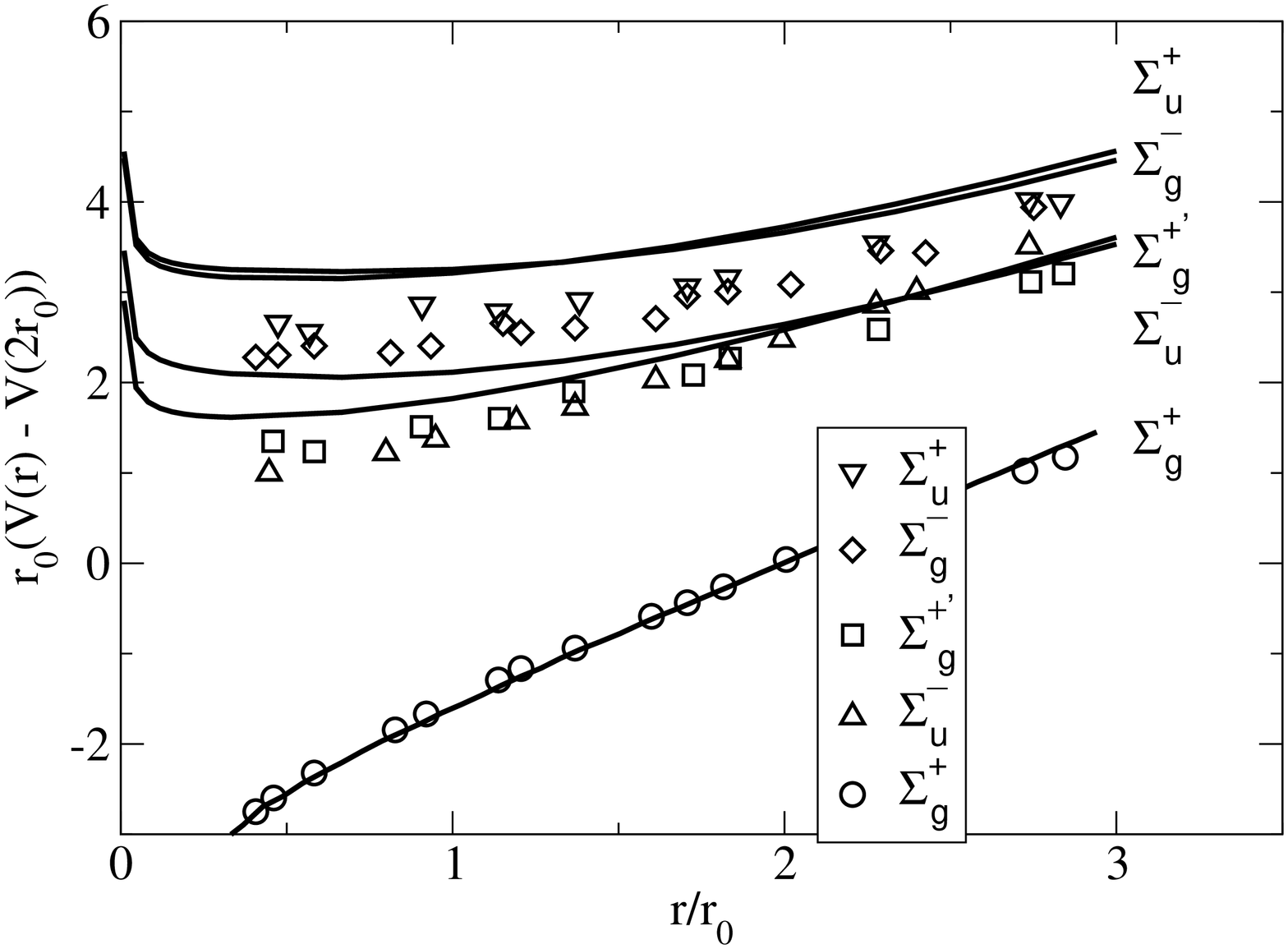}
\caption{\label{3b-s} Same as in Fig.~\ref{3b-p} for the $\Lambda=0$, $\Sigma$ energies. }
\end{figure}

\begin{figure}
\includegraphics[width=2.5in,angle=270]{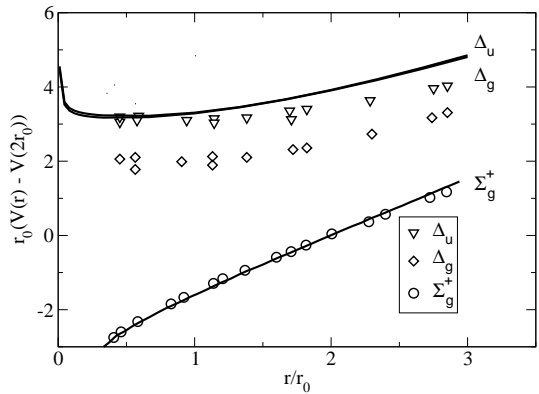}
\caption{\label{3b-d} Same as in Fig.~\ref{3b-p} for the $\Lambda=2$, $\Delta$ energies. }
 \end{figure}

To obtain the correct ordering it is necessary to include the irreducible, three-body interaction between quark, antiquark and gluon, shown in Fig.~\ref{me}e. This interaction is attractive and, as can be easily verified using the analytical expression in Eq.~(\ref{h3e}), as $R\to 0$ it pushes up in energy the $\xi =-1$ ($L_g = j_g$) states and has no effect on the $\xi=+1$ states. As seen in Figs.~\ref{3b-p},\ref{3b-s},\ref{3b-d} that show the results of diagonalization of the full Hamiltonian, this additional  interaction energy is sufficient to change the order of the $\Pi$ and excited $\Sigma$ states, and almost does the job 
 for the higher energy, $\Delta$ states.

\begin{figure}
\includegraphics[width=2.5in,angle=270]{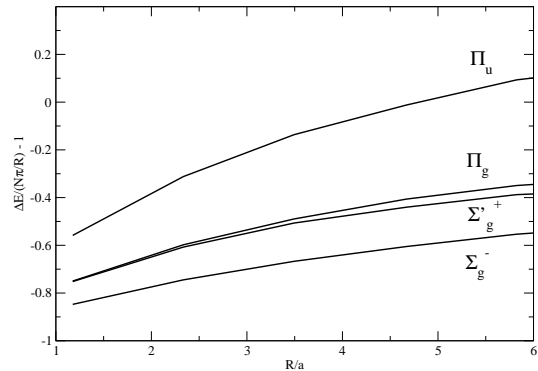}
\caption{\label{split} Splittings between the excited energies of quasi-gluons and the ground state  $Q{\bar Q}$ energy. The values $N$ expected from the string model are, ${\bf N=1}$ for $\Pi_u$, $N=2$ for $\Pi_g$ and $\Sigma'_g$ and $N=4$ for $\Sigma^-_g$ ($a=0.2\mbox{ fm}$). }
 \end{figure}

When studying the large $R$ behavior we first note that the long-range singularity of the linear potential is canceled  between the three self-energies and the two-body interactions. The  three body interaction between the quark, antiquark and gluon is IR finite, which also implies that in the limit $R\to \infty$ it decreases with $R$. This is due to the gradient coupling of the transverse gluon to the Coulomb line. Furthermore, connecting one gluon to the Coulomb line effectively {\it shortens} the Coulomb potential making it less confining~\cite{Szczepaniak:2005xi}.  All this implies that  splittings between excited states with $n_g+1$ and $n_g$ quasi-gluons at large $Q{\bar Q}$ separations do not increase with $R$ .  They are expected to be roughly constant corresponding  to  the average kinetic energy of a gluon in the color singlet state, $E_g \sim \omega(k\sim 1/R) = m_g \sim 600\mbox{MeV}$. On the other hand, to leading order in $1/R$, separations of string 
excitations are expected to be proportional to $\Delta E = E_{N+1} - E_N = N\pi/R$~\cite{Juge:2002br}.  In Fig.~\ref{split}  we plot $(\Delta E)/(N\pi/R) - 1$ with $\Delta E$ representing the energy difference between our $n_g=1$ excited quasi-gluon energies and the ground state $Q{\bar Q}$ energy. The corresponding values of $N$ were chosen as in~\cite{Juge:2002br}.  The 
 agreement with the lattice results, shown in Fig.~1 of ~\cite{Juge:2002br} is very good. In the string model 
 $(\Delta E)/(N\pi/R) - 1$ is expected to approach zero at large separations, while the lattice and our results seem to indicate a positive slope. As discussed above this slope can be interpreted  in terms of 
  quasi-particle excitations.

\section{Summary and Outlook} 
We computed the lowest excitation energies of the gluonic field in the presence of static $Q{\bar Q}$ sources. The gluons were described using single, quasi-particle states built above a mean-field variational ansatz for the ground state. The variational approach is expected to produce energy levels which are systematically higher than the true energies. Indeed the computed energies are above the lattice data by about one unit of $r_0^{-1}$, {\it i.e.} $400\mbox{ MeV}$. The nontrivial level ordering is however correctly reproduced once the  full effect of the Coulomb kernel is taken into account. In  particular the irreducible three-body force is the one which is responsible  of reversing the naive ordering expected in any constituent gluon model.  The quasi-gluon gluon description also reproduces the degeneracies in the energy levels for small separations, as well as the trends in the level splittings for intermediate separations of the order of a few fm. At larger separations, multi-gluon effects are expected and those  can be incorporated into this picture by renormalizing the bare $Q{\bar Q}$ potential. A quantitative analysis is currently being performed.

\section{Acknowledgment}
We would like to thank J.~Greensite and C.~Morningstar and 
 for several discussions and Scott Teige for his comments on the manuscript.  
  This work was supported in part by the US
Department of Energy grant under contract 
 DE-FG0287ER40365. 

\section{Appendix} 

For each $j_g$, the wave functions $\psi_{N,j_g}(k)$ are expanded in a complete  
  orthonormal basis of functions $\phi_{m,j_g}(k)$ 
  \begin{equation}
  \psi_{N,j_g}(k) = \sum_{m=1}^{m_{max}}  a^{m}_{N,j_g} \phi_{m,j_g}(k), 
  \end{equation} 
with normalization,  

\begin{equation} 
\int {{dk k^2} \over {(2\pi)^3}}   \phi^*_{m',j'_g}(k) \phi_{m,j_g}(k) = \delta_{m',m}\delta_{j'_g,j_g}.
\end{equation} 
  The expansion coefficients are computed by diagonalizing the 
   $(m_{max}j_{g,max})  \times (m_{max}j_{g,max})$  matrix, $\tilde{H}_{m'j_g';m,j_g}$,   obtained by evaluating the diagrams in Fig.~\ref{me} 

\begin{equation}
\tilde{H_3}  = H_{3a} + H_{3b} + \cdots + H_{3f}, \label{htot}
\end{equation} 
evaluated in the basis of functions $\phi_{m,j_g}$. In numerical computations  for each $j_g$, we used a momentum grid as the basis functions. 
    The numerical results presented were for a single $j_g$  determined from 
    Eq.~(\ref{PC}) after verifying that increasing $j_g$  changes the computed spectrum by at most a few percent. 
 For arbitrary $\Lambda^Y_{PC}$ the Hamiltonian matrix elements are given by 

\begin{equation} 
H_{3a} = {{\delta_{j'_g,j_g}}\over 2}  \int {{dk k^2} \over {(2\pi)^3}} \phi^*_{m',j_g}(k) E_g(k) \phi_{m,j_g}(k). 
\end{equation}
The single gluon energy, $E_g(k)$ is given in ~\cite{Szczepaniak:2001rg,Szczepaniak:2005xi}. 
\begin{eqnarray}
H_{3b} & = &  -  C_F V_C(0)\delta_{m',m}\delta_{j'_g,j_g}  \nonumber \\
 &  = &  - 4\pi C_F \int {{dk k^2} \over {(2\pi)^3}} V_C(k)  \delta_{m',m}\delta_{j'_g,j_g},
\end{eqnarray}
with $V_C(R)$ fitted to the ground state $Q{\bar Q}$ potential~\cite{Szczepaniak:2001rg,Szczepaniak:2005xi}. 
\begin{widetext}
\begin{eqnarray} 
H_{3c} & =  & 
{N_C \over 2} \sum_{\lambda,\lambda',\sigma,\sigma',\mu} \int {{d\q} \over {(2\pi)^3}} 
 \int {{d\k} \over {(2\pi)^3}} 
 \phi^*_{m',j'_g}(q) \phi_{m,j_g}(k)\int d\x  \left[ V_C(\x - { {\bf R}\over 2}) + V_C (\x + {{\bf R}\over 2}) \right]  
  e^{  i \x \cdot (\k-\q)  }   \nonumber \\
 & \times &   { \sqrt{ {(2j'_g + 1)(2j_g + 1)}} \over {16\pi} }
  \left[  D^{j'_g}_{\Lambda,\sigma'}(\hat\q) 
    D^{j_g,*}_{\Lambda\lambda'}(\hat\k) \chi^{\xi'}_{\sigma\sigma'} \chi^{\xi}_{\lambda\lambda'} 
    D^{1*}_{\mu\sigma}(\hat\q) D^1_{\mu\lambda}(\hat\k)  
     + \eta_Y\eta_Y' (\Lambda \to -\Lambda) \right] \left(\sqrt{ { \omega(k)} \over {\omega(q)}} 
   + \sqrt{ {\omega(q)} \over {\omega(k)}} \right)\nonumber \\
   & = & {N_C \over 2} 
\sum_{\lambda,\lambda',\sigma,\sigma',\mu} \int {{d\q} \over {(2\pi)^3}} \int {{d\k} \over {(2\pi)^3}} 
 \phi^*_{m',j'_g}(q) V_C(\k - \q)  \left[ e^{ - i {{\bf R} \over 2} \cdot (\k-\q)  } 
  +  e^{  i {{\bf R} \over 2} \cdot (\k-\q)  }  \right] \phi_{m,j_g}(k)
   \nonumber \\
 & \times &  
 {\sqrt{(2j'_g + 1)(2j_g + 1)} \over {16\pi} }   \left[  D^{j'_g}_{\Lambda,\sigma'}(\hat\q) 
    D^{j_g,*}_{\Lambda\lambda'}(\hat\k) \chi^{\xi'}_{\sigma\sigma'} \chi^{\xi}_{\lambda\lambda'} 
    D^{1*}_{\mu\sigma}(\hat\q) D^1_{\mu\lambda}(\hat\k)  
     + \eta_Y\eta_Y' (\Lambda \to -\Lambda) \right]  \left(\sqrt{ { \omega(k)} \over {\omega(q)}} 
   + \sqrt{ {\omega(q)} \over {\omega(k)}} \right), \nonumber \\
\end{eqnarray}
\end{widetext}
and $\eta_Y$ and $\xi$ related to $j_g$ and $\Lambda^Y_{PC}$  through  Eq.~(\ref{PC}). 

\begin{eqnarray}
H_{3d} & = &  -  {1\over {2N_C}} V_C(R)\delta_{m',m}\delta_{j'_g,j_g}  \nonumber \\
 &  = &  - 4\pi  {1\over {2N_C}}   \int {{dk k^2} \over {(2\pi)^3}} V_C(k) j_0(R k)\delta_{m',m}\delta_{j'_g,j_g} , \nonumber \\
 \end{eqnarray} 

\begin{widetext}
\begin{eqnarray}
H_{3e}  & = &   \sum \int {{d\k} \over {(2\pi)^3}} {{d\p} \over {(2\pi)^3}} {{d\q} \over {(2\pi)^3}} 
 {{\phi^*_{m',j'_g}(p)}\over {\sqrt{2\omega(p)}}} {{ \phi_{m,j_g}(k) }\over {\sqrt{2\omega(k)}}} 
  \nonumber \\ 
& \times &  \int d\x d\y d\z \left[  K(\x - {{\bf R}\over 2}, \z + \y - \x, \y + {{\bf R}\over 2})  + ({\bf R} \to - {\bf R}) \right] 
 e^{i\x \cdot \k} e^{i \z \cdot \q} e^{-i \y \cdot \p} \nonumber \\
 & \times &   { \sqrt{ {(2j'_g + 1)(2j_g + 1)}} \over {8\pi} } \left[
  D^{j'_g}_{\Lambda,\sigma'}(\hat\p) D^{1,*}_{\mu,\sigma}(\hat\p) \chi^{\xi'}_{\sigma'\sigma}
 D^1_{\mu,0}(\hat\q) 
  D^{j_g,*}_{\Lambda,\lambda'}(\hat\k) D^1_{\nu,\lambda}(\hat\k) \chi^{\xi}_{\lambda'\lambda}
 D^{1,*}_{\nu,0}(\hat\q)  + \eta_Y\eta'_Y (\Lambda \to -\Lambda) \right] \nonumber \\
 & = & 
 \sum \int {{d\k} \over {(2\pi)^3}} {{d\p} \over {(2\pi)^3}} {{d\q} \over {(2\pi)^3}} 
{{ \phi^*_{m',j'_g}(p)}\over {\sqrt{2\omega(p)}}}  {{\phi_{m,j_g}(k) }\over {\sqrt{2\omega(k)}}} K(\k+\q,\q,\p+\q)  
\left[  e^{i {{\bf R}\over 2} \cdot ( \k + \p + 2\q)} + ({\bf R} \to -{\bf R}) \right] 
\nonumber \\
 & \times &   { \sqrt{ {(2j'_g + 1)(2j_g + 1)}} \over {8\pi} } \left[
  D^{j'_g}_{\Lambda,\sigma'}(\hat\p) D^{1,*}_{\mu,\sigma}(\hat\p) \chi^{\xi'}_{\sigma'\sigma}
 D^1_{\mu,0}(\hat\q) 
  D^{j_g,*}_{\Lambda,\lambda'}(\hat\k) D^1_{\nu,\lambda}(\hat\k) \chi^{\xi}_{\lambda'\lambda}
 D^{1,*}_{\nu,0}(\hat\q)  + \eta_Y\eta'_Y (\Lambda \to -\Lambda) \right], \nonumber \\ \label{h3e}
 \end{eqnarray}
 \end{widetext}
 where the sum is over $\mu,\nu,\lambda,\lambda',\sigma,\sigma'$ and the kernel is given by 
 \begin{equation}
K(\x,\z,\y) 
 =   \int {{d\k} \over {(2\pi)^3}} {{d\p} \over {(2\pi)^3}} {{d\q} \over {(2\pi)^3}}
  K(k,q,p)  e^{i\x\cdot \k} e^{i\y \cdot \p} e^{i\z\cdot \q},   
 \end{equation}
 and the kernel $K(k,q,p)$ is given in ~\cite{Szczepaniak:2001rg,Szczepaniak:2005xi}.
 
Finally, 
 \begin{widetext}
\begin{eqnarray}
H_{3f}  & = &   \sum \int {{d\k} \over {(2\pi)^3}} {{d\p} \over {(2\pi)^3}} {{d\q} \over {(2\pi)^3}} 
 {{\phi^*_{m',j'_g}(p)}\over {\sqrt{2\omega(p)}}} {{ \phi_{m,j_g}(k)}\over {\sqrt{2\omega(k)}}} 
 \nonumber \\
& \times &   \int d\x d\y d\z \left[  K(\x - {{\bf R}\over 2}, \z + \y - \x, \y - {{\bf R}\over 2})  + ({\bf R} \to - {\bf R}) \right] 
 e^{i\x \cdot \k} e^{i \z \cdot \q} e^{-i \y \cdot \p} \nonumber \\
 & \times &   { \sqrt{ {(2j'_g + 1)(2j_g + 1)}} \over {8\pi} } \left[
  D^{j'_g}_{\Lambda,\sigma'}(\hat\p) D^{1,*}_{\mu,\sigma}(\hat\p) \chi^{\xi'}_{\sigma'\sigma}
 D^1_{\mu,0}(\hat\q) 
  D^{j_g,*}_{\Lambda,\lambda'}(\hat\k) D^1_{\nu,\lambda}(\hat\k) \chi^{\xi}_{\lambda'\lambda}
 D^{1,*}_{\nu,0}(\hat\q)  + \eta_Y\eta'_Y (\Lambda \to -\Lambda) \right] \nonumber \\
 & = & 
 \sum \int {{d\k} \over {(2\pi)^3}} {{d\p} \over {(2\pi)^3}} {{d\q} \over {(2\pi)^3}} 
{{ \phi^*_{m',j'_g}(p)}\over {\sqrt{2\omega(p)}}}{{ \phi_{m,j_g}(k)}\over {\sqrt{2\omega(k)}}}   K(\k+\q,\q,\p+\q)  
\left[  e^{i {{\bf R}\over 2} \cdot ( \k - \p )} + ({\bf R} \to -{\bf R}) \right] 
\nonumber \\
 & \times &   { \sqrt{ {(2j'_g + 1)(2j_g + 1)}} \over {8\pi} } \left[
  D^{j'_g}_{\Lambda,\sigma'}(\hat\p) D^{1,*}_{\mu,\sigma}(\hat\p) \chi^{\xi'}_{\sigma'\sigma}
 D^1_{\mu,0}(\hat\q) 
  D^{j_g,*}_{\Lambda,\lambda'}(\hat\k) D^1_{\nu,\lambda}(\hat\k) \chi^{\xi}_{\lambda'\lambda}
 D^{1,*}_{\nu,0}(\hat\q)  + \eta_Y\eta'_Y (\Lambda \to -\Lambda) \right]. \nonumber \\
 \end{eqnarray}
 \end{widetext}

\end{document}